\documentstyle[11pt,paspconf,epsf]{article}

\begin{document}

\title{Radiation-Pressure Ejection of Planetary Nebulae
in Asymptotic-Giant-Branch Stars}
\author{A. V. Sweigart}
\affil{NASA Goddard Space Flight Center, Code 681, Greenbelt, MD 20771}

\begin{abstract}
We have investigated the possibility that radiation pressure
might trigger planetary nebula (PN) ejection during helium-shell
flashes in asymptotic-giant-branch (AGB) stars.  We find that the
outward flux at the base of the hydrogen envelope during a flash
will reach the Eddington limit when the envelope mass $M_{\rm env}$
falls below a critical value that depends on the core mass $M_{\rm H}$
and composition.  These results may help to explain
the helium-burning PN nuclei found in the
Magellanic Clouds.
\end{abstract}


\section{Description of Radiation-Pressure Instability}

We have computed extensive AGB evolutionary sequences for 
heavy-element abundances Z = 0.01716 (solar) and 0.002 in order to
study a radiation-pressure instability for ejecting a PN during
a helium-shell flash.  The existence of such an instability
was previously confirmed by Wood \& Faulkner (1986) but only
for more massive AGB stars with large core masses
($M_{\rm H} \, > \, 0.86 \, M_{\sun}$).  In contrast, we find this
instability at core masses as small as $M_{\rm H} \, \approx \, 0.6 \, M_{\sun}$
due, in part, to our use of the new OPAL opacities
(Rogers \& Iglesias 1992).

Results for a typical sequence are presented in the left panels
of Figure 1.  The strong flashes in this sequence cause
the hydrogen shell to expand outward to very low
temperatures and densities.  This, in turn, lowers the gas
pressure and hence the minimum value $\beta_{\rm min}$ of the
ratio $\beta$ of gas to total pressure near the base of the
hydrogen envelope.  Most importantly, we note from Figure 1 that
$\beta_{\rm min}$ decreases monotonically from flash to
flash and eventually goes to 0, implying that the entire envelope
is then supported by radiation pressure.

During the final flashes in Figure 1 the temperature within the
hydrogen shell falls to log $T$ = 5.3 at which there is a well-known
peak in the OPAL opacities.  When the shell reaches this peak,
its opacity suddenly increases, and, as a result, its Eddington
luminosity abruptly drops.  This occurs at a time when the
outward flux within the shell is considerably enhanced by the
flash.  As the shell continues to cool, its opacity increases further,
thereby forcing an even greater expansion and driving $\beta$ down
to even smaller values.  The star thus encounters an ``opacity catastrophe''
whose likely outcome is envelope ejection.

The present sequences show that there is a critical $M_{\rm env}$
at which radiation pressure will support the hydrogen envelope
during a flash (see right panel of Figure 1).  This critical
$M_{\rm env}$ increases linearly with $M_{\rm H}$ and decreases
with decreasing metallicity, since the peak in the OPAL opacities
is then smaller.  The critical values of $M_{\rm env}$ in Figure 1
represent the minimum envelope mass at
\newpage

\noindent
\hskip -0.85in
\parbox{3.5in}{\epsfxsize=3.5in \epsfbox{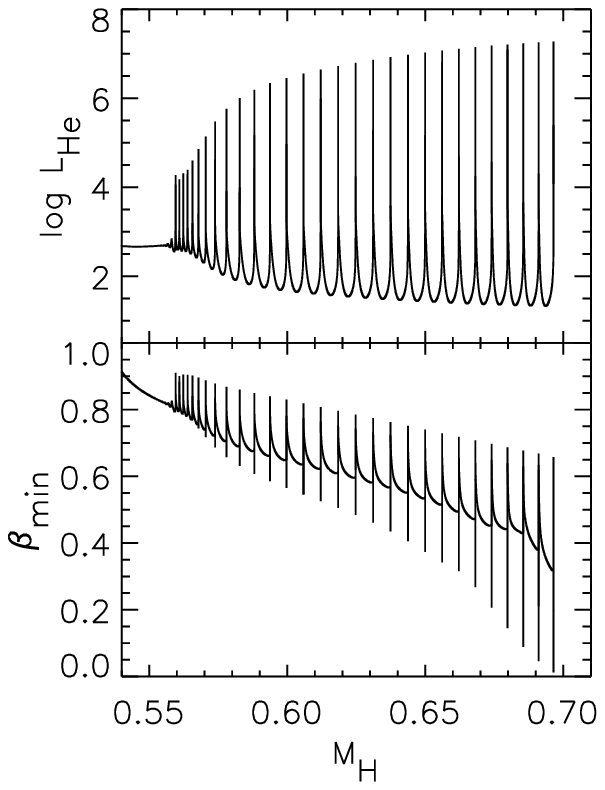}}
\hskip -0.83in
\parbox{3.43in}{\epsfxsize=3.43in \epsfbox{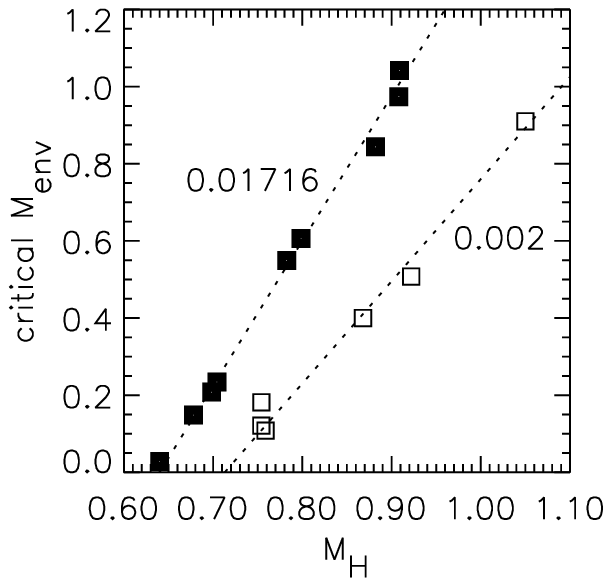}}

\begin{center}
\parbox{4.75in}{Figure 1.  Helium-burning luminosity
$L_{\rm He}$ in solar units (upper left panel) and minimum
value $\beta_{\rm min}$ of the ratio $\beta$ of gas to total pressure
near the base of the hydrogen envelope (lower left panel) as a function
of the core mass $M_{\rm H}$ during the helium-shell flashes
of a solar metallicity AGB star.  The right panel shows the dependence
of the critical envelope mass $M_{\rm env}$ at which $\beta$ goes to 0
during a flash on $M_{\rm H}$ for two heavy-element abundances:
Z = 0.01716 (solid squares) and 0.002 (open squares).  The dashed
lines are linear fits to the model data.}
\end{center}

\vskip \baselineskip 
\vskip \baselineskip 
\noindent
which an AGB star
can undergo a flash without $\beta$ going to 0.  These
results depend, however, on the extent of 3rd dredge-up
in the model calculations.

The minimum in $\beta$ during a flash occurs at the
time when 3rd dredge-up is most likely.  We suggest that such
dredge-up might trigger PN ejection by increasing the opacity
and thereby lowering the Eddington luminosity at the base
of the envelope, especially at low metallicites where
dredge-up is favored by current theoretical models.  This
might explain why the majority of halo PN's and most non-Type I
PN's in the Magellanic Clouds are carbon rich.  It might
also explain why the PN K648 in M15 has a high carbon
abundance despite the lack of AGB carbon stars in this globular
cluster.  We suggest therefore that the high carbon abundance
of K648 was the cause of its PN ejection.

\end{document}